# Nuclear Georeactor Generation of Earth's Geomagnetic Field

by

J. Marvin Herndon
Transdyne Corporation
San Diego, CA 92131 USA

July 19, 2007

**Abstract:** The purpose of this communication is to suggest that the mechanism for generating the geomagnetic field and the energy source for powering it are one and the same, a nuclear georeactor at the center of the Earth. Toward that end, I : *i*) Present evidence that the nuclear georeactor fission-product sub-shell is fluid; *ii*) Suggest that the geomagnetic field is generated within the georeactor sub-shell, rather than within Earth's iron-alloy fluid core; *iii*) Describe why convection appears more feasible within the georeactor sub-shell than within the iron-alloy core; *iv*) Disclose additional relative physical advantages for georeactor sub-shell dynamo operation; and, *v*) Outline briefly the research that should be conducted to advance the state of knowledge of georeactor-geomagnetic field generation. The concept of geomagnetic field production by the nuclear georeactor is presented specifically for the Earth. The concepts and principles, however, are generally applicable to planetary magnetic field production.

No other manifestation of Earth has been as seemingly inexplicable as the Earth's magnetic field. More than a thousand years ago, individuals in China set afloat in bowls of water tiny slivers of loadstone, the mineral now called magnetite, and discovered that the slivers quickly assumed a preferred direction. That observation led to the development of the magnetic compass, which is still in use by navigators, hikers, and others seeking a way to determine direction. William Gilbert's *Die Magnete* (1600), his definitive work, was based upon extensive magnetic measurements collected around the globe, which showed that the Earth itself is like a giant magnet, rather than the magnetism arising from an extraterrestrial source as supposed by others [1]. In 1838, the mathematical genius, Johann Carl Friedrich Gauss, proved that the Earth's magnetism source is at, or very near, the center of the Earth [2].

The Earth acts like a giant magnet with a magnetic field extending into interplanetary space, shielding the planet by deflecting charged particles of the solar wind, but it is not a permanent magnet. The Earth's magnetic field must continuously be fed with energy; otherwise its interactions with matter in the Earth would cause it to decay and eventually disappear. Thus, there must exist at or near the Earth's center a mechanism for generating the geomagnetic field and an energy source that powers it. The purpose of this paper is to suggest that the mechanism for generating the geomagnetic field and the energy source for powering it are one and the same, a nuclear georeactor at the center of the Earth.

In 1919, Larmor suggested that the Sun's magnetic field might be sustained by a mechanism similar to a self-exciting dynamo. Elsasser [3-5] and Bullard [6] first adapted the solar dynamo concept to explain the generation of the Earth's magnetic field. Decades of subsequent research efforts have focused almost exclusively along those lines of thought, namely, that the geomagnetic field is believed to originate by convective motions in the Earth's fluid core interacting with Coriolis forces produced by planetary rotation [7]. Reference is frequently made to the great complexities involved in the calculations, but little thought seems to have been given to the possibility that there might be fundamental errors in the underlying assumptions, especially in the assumption that the geomagnetic field originates in the Earth's fluid core.

Experimentalists are keenly aware of propagation and persistence of errors in calculations. But the concept of error propagation and persistence may apply in a more general sense to the adverse consequence of uncorrected error in scientific understanding. The matter that is the Earth is connected in fundamental ways to the matter of certain chondrite meteorites, and to the matter of the outer portion of the Sun [8, 9]. There is good reason to believe that the Earth is, in the main, like a chondrite meteorite. Beginning as early as 1940, scientists, including Francis Birch, built geophysics upon the premise that the Earth is like ordinary chondrites, the most common type of meteorite observed impacting Earth, while totally ignoring another, albeit less abundant type, called enstatite chondrites. The principal difference between the two meteorite types is that enstatite chondrites formed under circumstances of extremely limited available oxygen, leading to certain normally oxyphile elements existing either partially or wholly in the alloy portion that corresponds to the core of the Earth.

Beginning in 1979, I published a series of papers revealing a fundamentally different view of Earth's interior, including the inner core being nickel silicide, Earth-core precipitates CaS and MgS at the core mantle boundary, the lower mantle consisting of essentially FeO-free $MgSiO_3$, and the boundary between the upper and lower mantle being a compositional boundary with the matter below that boundary, the endo-Earth, being like an enstatite chondrite [8-11]. Those discoveries and insights led to a fundamentally different view of Earth formation, dynamics, energy production, and energy transport process [12-14].

In 1993, I published the first of a series of papers revealing the background, feasibility and evidence of a nuclear fission georeactor at the center of the Earth as the energy source for the geomagnetic field [14-19], which, as Rao notes with extensive references [20], may offer the solution to the riddles of geomagnetic field variability and deep-Earth helium production. The calculations underlying my georeactor concept have been verified [21] and extensive numerical simulations have been conducted at Oak Ridge National Laboratory [19, 22] and using computer software licensed there from [14, 23].

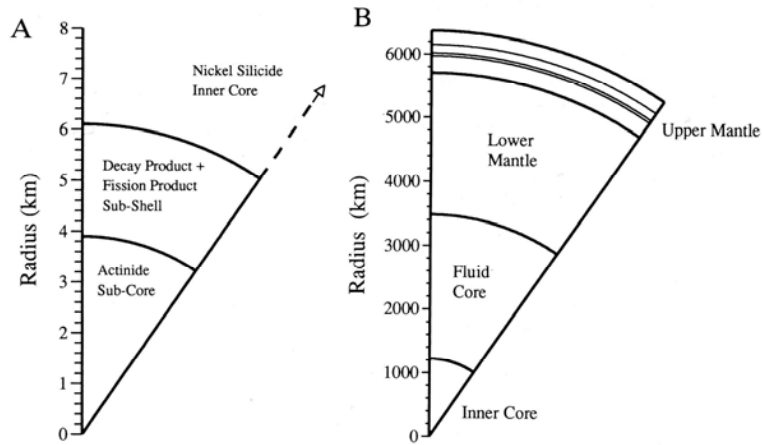

**Figure 1.** Schematic representation of georeactor sub-structure within the Earth's inner core (A) and (B) internal structure of Earth as a whole. Adapted from reference [17].

Briefly, the underlying basis of the nuclear georeactor, shown schematically in Figure 1, is as follows: Seismologically-determined parts of the interior of the Earth are related through fundamental ratios of mass to corresponding, mineralogically-determined parts of the Abee enstatite chondrite. By that identity, the observation that uranium occurs in the Abee meteorite, not in its silicate, but in its alloy components implies the existence within the Earth's core of as much as 82% of our planets uranium content. Uranium, being incompatible in an iron-based alloy, is expected to precipitate at a high temperature, perhaps as the compound US. As density at Earth-core pressures is a function almost exclusively of atomic number and atomic mass, uranium, or a compound thereof, would be the core's most dense precipitate and would tend to settle, either directly or through a series of steps, by gravity to the center of the Earth, where it would quickly form a critical mass and become capable of self-sustained nuclear fission chain reactions. Georeactor numerical simulations indicate that the georeactor would function as a fast neutron breeder reactor capable of operating for at least as long as Earth has existed. Georeactor operation produces energy and fission products. The fission products, having atomic numbers and atomic masses approximate half that of the actinide fuel, are expected to migrate radially outward, forming a fission-product sub-shell surrounding the actinide reactor sub-core [17]. The whole georeactor assembly is expected to exist at the center of Earth in contact with and surrounded by the nickel silicide inner core.

The purposes of this communication are the following: *i*) To present evidence that the nuclear georeactor fission-product sub-shell is fluid; *ii*) To suggest that the geomagnetic field is generated within the georeactor sub-shell, rather that within Earth's iron-alloy fluid core; *iii*) To describe why convection appears more feasible within the georeactor sub-shell than within the iron-alloy core; *iv*) To disclose additional relative physical advantages for georeactor sub-shell dynamo operation; and, *v*) To outline briefly the research that should be conducted to advance the state of knowledge of georeactor-geomagnetic field generation.

In 1996, I estimated the present radius of the georeactor actinide sub-core to be 4 km and the outer boundary of the fission-product sub-shell to be about 6 km. These were very conservative estimates, perhaps reasonable lower limits, within the unknowns; the actual dimensions may be as much as several times greater. In addition to fission products, the sub-shell may contain appreciable lead, which should be liquid, from the decay of actinide elements. Lead has quite a low melting point; at Earth-core pressures one would expect lead to be molten at temperatures well below the temperature at which an iron-sulfur alloy, such as the core, would be fluid.

When an actinide nucleus fissions, it typically splits into two pieces, each consisting of one out of about 1000 possible fission product nuclides, most radioactive, and most with a very short half-life. Table 1 shows the cumulative fission yield fractions, expressed as elements, for the most abundant elements produced by fast neutron fission of $^{238}$U and $^{235}$U. The values tabulated represent fission-produced elements whose fraction is >0.0001 after transmutation of very short half-life precursors. In the absence of data on the melting point of that mix of elements at Earth-core pressures, I plot in Figure 2 the fractional amount of each of those element fractions and the melting point of the corresponding, pure element at ambient pressure. As a reference standard, the dashed vertical line shows the melting point of nickel silicide, structure Ni$_2$Si, at ambient pressure.

At ambient pressure nearly 60% of the fission-produced elements shown in Figure 2 have melting points considerably less than nickel silicide. That suggests, although does not prove, that the georeactor sub-shell at the center of Earth is fluid. Further considerations lend support to potential fluidity. Generally, in at least the vast majority of cases at ambient pressure, mixtures of elements tend to have melting points lower than their higher melting point components. Moreover, the mix of fission-produced element-fractions shown in Figure 2 does not include natural-decay-produced lead, with its low ambient pressure melting point of 328°C. Furthermore, it should be noted that even the presence of up to about 50% solids, as slurry, will still behave as a liquid. For these reasons, the nuclear georeactor fission-product sub-shell is believed to be fluid.

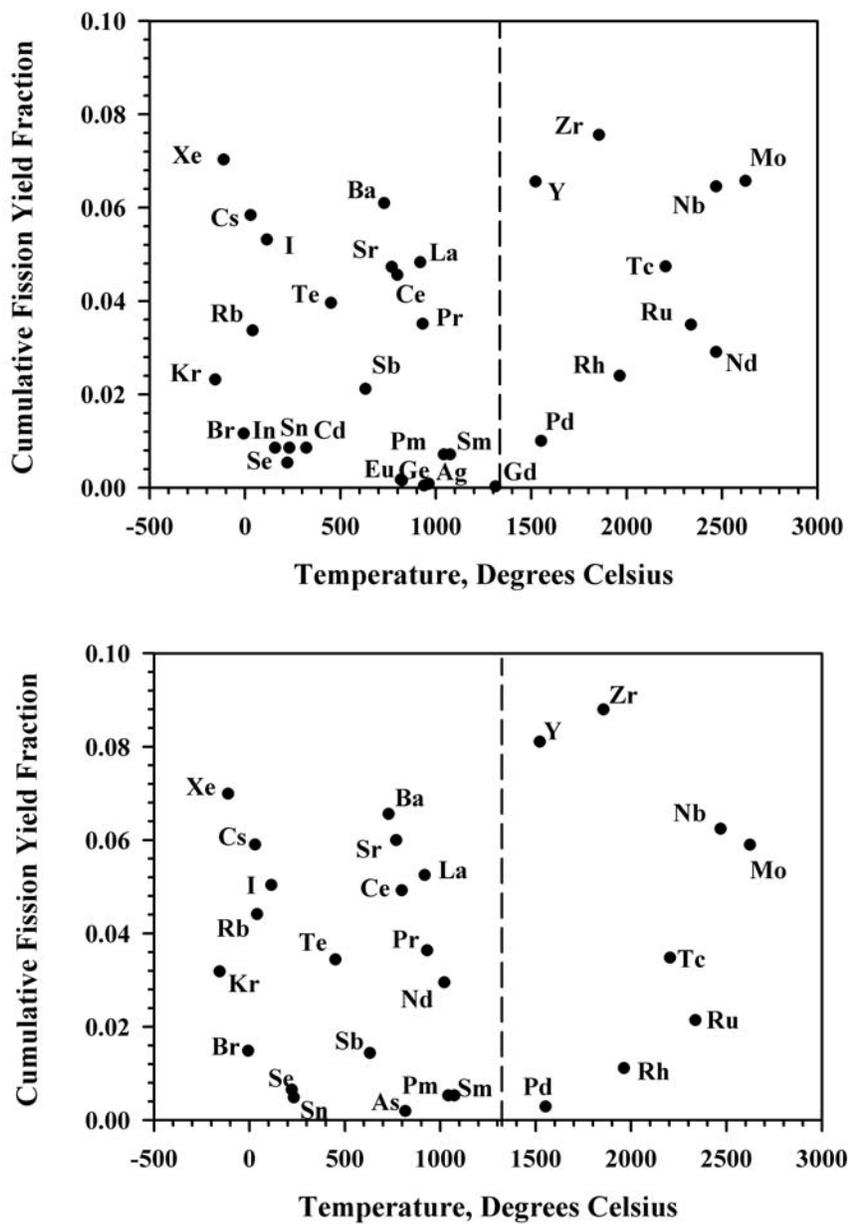

**Figure 2.** Cumulative yield fraction of $^{238}$U and $^{235}$U fast neutron fission product elements plotted vs. ambient pressure melting point of each respective element. The dashed vertical line represents the ambient pressure melting temperature of nickel silicide, $Ni_2Si$.

The much missed Subramanyan Chandrasekhar (1910-1995) defines convection in the following way [24]: "The simplest example of thermally induced convection arises when a horizontal layer of fluid is heated from below and an adverse temperature gradient is maintained. The adjective 'adverse' is used to qualify the prevailing temperature gradient, since, on account of thermal expansion, the fluid at the bottom becomes lighter than the fluid at the top; and this is a top-heavy arrangement which is potentially unstable. Under these circumstances the fluid will try to redistribute itself to redress this weakness in its arrangement. This is how thermal convection originates: It represents the efforts of the fluid to restore to itself some degree of stability."

Even with heat being supplied by the georeactor, and conducted through the nickel silicide inner core to the base of the fluid core, a fundamental difficulty arises in maintaining an adverse temperature gradient in the iron-sulfur core. To maintain an adverse temperature gradient in the core over extended time, the heat continuously brought to the top of the core by convection must be continuously removed at the same rate. But the silicate mantle above the core-interface is much more of an insulator than a thermal conductor. The core is thermally well insulated. That difficulty does not arise with convection in the nuclear georeactor sub-core region. Nuclear fission produced heat is supplied directly to the base of the fission-product sub-shell. The outer boundary of the fluid sub-shell maintains contact with the semi-metallic, nickel silicide inner core, which acts as a heat sink, a thermal ballast, with reasonably good thermal conductivity to transport excess heat to the fluid iron-sulfur core. This arrangement enables the sub-shell's fluid to restore to itself, and to maintain, a reasonably good degree of stability. Convection stability in the sub-shell is also enhanced by the lower gravitational acceleration and by the considerably shorter scale length than exists in the iron-sulfur core.

The dynamo mechanism, thought to be responsible for generating the geomagnetic field, operates as a magnetic amplifier wherein, beginning with a small magnetic field, the combined motions of an electrically conducting fluid, driven by convection in a rotating system, amplify and maintain a more-or-less stable, much, much larger magnetic field. The absence of a seed-field is another major objection to the idea of geomagnetic field production within the iron-sulfur core. By contrast, that problem is wholly obviated with nuclear georeactor sub-shell magnetic field production. Heavy elements, like uranium, are relatively more neutron-rich than lighter elements. Consequently, when uranium fissions, not only are neutrons liberated, which are necessary for maintaining a nuclear chain reaction, but the fission products formed are themselves neutron-rich. Neutron-rich fission products are frequently electrically conducting, generally radioactive, and usually decay by the emission of a beta particle, a negatively charged electron. Beta decay, as well as other ionizing radiation, may provide ample opportunity for the formation of the requisite dynamo seed-field produced by the motion of charged particles or by an electrical current arising from the separation of charges.

The concept of geomagnetic field production by the dynamo mechanism involving convection in the nuclear georeactor fission-product sub-shell, driven by nuclear fission energy produced in the georeactor sub-core, is fundamentally different and more efficacious than the postulated iron-sulfur core dynamo first suggested by Elsasser [3-5] and Bullard [6], which has been the subject of a plethora of investigations over a period of more than a half century. But for the georeactor-geodynamo theory, what are sorely needed are experimental investigations of the physical properties of the relevant materials at deep-Earth pressures and temperatures. Although I first suggested the idea of the inner core being composed of nickel silicide in 1979, to date no one has published experimental data on the stable crystal structures and compositions or on physical properties, such as melt relations, thermal conductivity, or on its electrical properties under deep-Earth conditions. These should be done, without delay. In addition, considerations should be given to the aggregate behavior of fission-product mixtures.

The concept of geomagnetic field production by the nuclear georeactor is presented above specifically for the Earth. The concepts and principles, however, are generally applicable to planetary magnetic field production.

## Acknowledgements

I have benefitted from discussions with and appreciate the encouragement from Maj. Dr. David Byers, Dr. P. K. Iyengar, Dr. Joseph Magill, and Dr. W. Seifritz.

Table 1. Cumulative fission yield fractions for fast neutron fission of $^{238}$U and $^{235}$U from tabulations posted on http://www.nucleonica.net

| Element | $^{238}$U Fission Fraction | $^{235}$U Fission Fraction | Element | $^{238}$U Fission Fraction | $^{235}$U Fission Fraction |
|---|---|---|---|---|---|
| Zirconium | 0.0756 | 0.0880 | Yttrium | 0.0656 | 0.0811 |
| Xenon | 0.0703 | 0.0699 | Barium | 0.0610 | 0.0656 |
| Niobium | 0.0645 | 0.0624 | Strontium | 0.0473 | 0.0600 |
| Cesium | 0.0456 | 0.0590 | Molybdenum | 0.0658 | 0.0590 |
| Lanthanum | 0.0483 | 0.0525 | Iodine | 0.0532 | 0.0504 |
| Cerium | 0.0492 | 0.0492 | Rubidium | 0.0337 | 0.0442 |
| Praseodymium | 0.0351 | 0.0364 | Technetium | 0.0474 | 0.0348 |
| Tellurium | 0.0396 | 0.0344 | Krypton | 0.0231 | 0.0318 |
| Neodymium | 0.0291 | 0.0295 | Ruthenium | 0.0349 | 0.0215 |
| Bromine | 0.0116 | 0.0148 | Antimony | 0.0212 | 0.0144 |
| Rhodium | 0.0240 | 0.0111 | Selenium | 0.00545 | 0.0065 |
| Samarium | 0.0071 | 0.0053 | Promethium | 0.0071 | 0.0053 |
| Tin | 0.0085 | 0.0048 | Palladium | 0.0100 | 0.0029 |
| Arsenic | 0.0018 | 0.0020 | Europium | 0.0015 | 0.0008 |
| Indium | 0.0085 | 0.0007 | Silver | 0.0008 | 0.0006 |
| Cadmium | 0.0085 | 0.0005 | Germanium | 0.0005 | 0.0005 |
| Gadolinium | 0.0003 | 0.0001 | Gallium | 0.0001 | 0.0001 |